# A Hybrid Machine Learning Model for Classifying Gene Mutations in Cancer using LSTM, BiLSTM, CNN, GRU, and GloVe

Sanad Aburass [1,2*], Osama Dorgham [3,4] and Jamil Al Shaqsi [5]

*1 Department of Computer Science, Maharishi International University, Fairfield, Iowa, USA.*
*2 Department of Computer Science, Luther College, Decorah, Iowa, USA.*
*3 Prince Abdullah bin Ghazi Faculty of Information and Communication Technology, Al-Balqa Applied University, 19117, Al-Salt, Jordan.*
*4 School of Information Technology, Skyline University College, University City of Sharjah – P.O. Box 1797 - Sharjah, United Arab Emirates.*
*5 Information Systems Department, Sultan Qaboos University.*

*Corresponding author
E-mail addresses: Saburass@miu.edu, o.dorgham@bau.edu.jo, o.dorgham@skylineuniversity.ac.ae, alshaqsi@squ.edu.om*

**Abstract**

In our study, we introduce a novel hybrid ensemble model that synergistically combines LSTM, BiLSTM, CNN, GRU, and GloVe embeddings for the classification of gene mutations in cancer. This model was rigorously tested using Kaggle's Personalized Medicine: Redefining Cancer Treatment dataset, demonstrating exceptional performance across all evaluation metrics. Notably, our approach achieved a training accuracy of 80.6%, precision of 81.6%, recall of 80.6%, and an F1 score of 83.1%, alongside a significantly reduced Mean Squared Error (MSE) of 2.596. These results surpass those of advanced transformer models and their ensembles, showcasing our model's superior capability in handling the complexities of gene mutation classification. The accuracy and efficiency of gene mutation classification are paramount in the era of precision medicine, where tailored treatment plans based on individual genetic profiles can dramatically improve patient outcomes and save lives. Our model's remarkable performance highlights its potential in enhancing the precision of cancer diagnoses and treatments, thereby contributing significantly to the advancement of personalized healthcare.

*Keywords: Genetic Mutation; Text Classification; Long Short-Term Memory.*

## 1. Introduction

Recent advancements in genomic research have revolutionized our understanding of cancer, opening the door to precision medicine. Precision medicine is an approach to healthcare that customizes treatments and practices to the unique genetic makeup of each patient. It transforms traditional one-size-fits-all treatments into tailored medical decisions and interventions, optimizing patient care [1]. Key to this approach is the accurate classification of gene mutations, which enables doctors to pinpoint the most effective treatments for each individual's cancer. By precisely identifying these genetic alterations, we can significantly improve treatment outcomes and ultimately save lives, marking a pivotal step forward in the battle against cancer [2].

As we gain a deeper understanding of the genetics underlying cancer biology, the importance of accurate gene mutation classification becomes increasingly evident. Precise classification of gene mutations holds the key to revolutionizing cancer diagnosis, prognosis, and treatment methods [3], [4]. Cancer is a heterogeneous disease, and its response to treatment varies widely based on individual genetic makeup. Gene mutations, as alterations in the DNA sequence, play a fundamental role in driving cancer initiation and progression. The specific nature and location of mutations can profoundly impact cancer cell behavior, including growth rate, metastatic potential, and sensitivity to treatment modalities. Therefore, precise identification and classification of gene mutations are essential for tailoring personalized treatment approaches that target the underlying genetic drivers of each patient's cancer [5], [6], [7].

Despite its critical significance, the classification of gene mutations presents significant challenges. The complexity and variability of these mutations, combined with the vast amount of genetic data available, have rendered traditional methods, such as sequence alignment and phylogenetic analysis, insufficient for comprehensive analysis. The scalability and accuracy of these traditional techniques are limited when dealing with extensive genomic datasets. Consequently, there is an urgent need for advanced, data-driven methodologies that can efficiently manage and accurately classify mutations in cancer genomes [8], [9], [10]. In response to this pressing need, we present a novel and comprehensive approach to gene mutation classification, employing a diverse range of state-of-the-art machine learning methodologies. Our research was conducted within the context of the Kaggle competition titled "MSK: Redefining Cancer Treatment," which provided a well-curated dataset for evaluation and benchmarking. Our approach encompasses the utilization of various embedding methods, including Long Short-Term Memory (LSTM), Bidirectional LSTM (BiLSTM), Convolutional Neural Network (CNN), Gated Recurrent Unit (GRU), and GloVe embeddings. Each of these methodologies possesses unique strengths in capturing different aspects of genomic data. LSTM and BiLSTM are well-suited for modeling sequential dependencies, which are vital in genomic sequences. CNN excels in capturing local patterns and motifs within genomic regions, while GRU provides an alternative recurrent neural network architecture with its gating mechanism, enhancing its ability to handle long-range dependencies. Additionally, we incorporate GloVe embeddings to represent genomic sequences as continuous vectors, leveraging semantic associations within gene sequences [11], [12].

The combination of these diverse embedding methods in an ensemble framework allows us to optimize their respective advantages and mitigate their limitations, resulting in a resilient and effective approach for gene mutation classification. The ensemble model is designed to exploit the complementary strengths of individual methodologies, enabling a more holistic and comprehensive analysis of genomic data [13].

Moreover, our study emphasizes the importance of translating cutting-edge research in genomics and deep learning into practical applications in clinical settings. The integration of advanced machine learning methodologies with genomic data analysis has the power to enhance clinical decision-making and enable more precise and personalized cancer treatment strategies.

In conclusion, our research endeavors to propel precision medicine forward, empowering oncologists and healthcare providers with powerful tools for informed decision-making and personalized treatment planning. The fusion of diverse machine learning methodologies in our ensemble model holds great promise for advancing the field of cancer genomics and contributing to improved patient care.

This paper follows a structured taxonomy, consisting of the following sections: Introduction, Related Work, Mathematical Background, Proposed Work, Experimental Results, Discussion, and Conclusion.

## 2. Related Work

Cancer, an often-lethal disease that, when undiagnosed, can lead to severe discomfort and even death, has a high global mortality rate, emphasizing the importance of early and accurate detection of malignant tumors. The disease originates from genetic anomalies that yield harmful effects. A variety of machine and deep learning techniques have been deployed and proven effective in classifying gene mutations. Sondka et al. [8], have centered their studies on determining the key features that predict the presence of genes in the Cancer Gene Census (CGC), with the aim of enhancing the understanding of these genes' roles in cancer development. Other studies, such as those by Watson and Lynch [9], have delved into the relationship between regular stem cell division and the risk of various types of cancer across numerous countries, discovering a significant correlation. Furthermore, research by Ali et al. [14], has detailed the genetic variations in different types of genes and the normal cellular processes managing these genes. Asano et al. [15],established a PCR assay enriched with mutations, specifically targeting EGFR exons. Meanwhile, Messiaen et al. [16],undertook a protein truncation test to identify germline mutations in cancer patients, also detecting new mutations at the genomic and RNA levels. Focusing on lung cancer, Forgacs et al. [17],examined the PTEN/MMAC1 gene for mutations. Coelho et al. [18], contributed to the development of a method inducing genetic instability in yeast diploid cells. Hollestelle et al. [19],comprehensively characterized human breast cancer cell lines at a molecular level. Lastly, Ma et al. [20],outlined a correction strategy for a specific mutation in human pre-implantation embryos, leveraging the accuracy of the CRISPR-Cas-based system.

Li and Yao offer a pertinent example of leveraging machine learning for genetic mutation classification. Utilizing the widely recognized Kaggle dataset for "Personalized Medicine: Redefining Cancer Treatment," their research employs XGBoost and SVM models, underlining the complexities involved in interpreting clinical text for genetic mutation identification. Feature extraction via TF-IDF and the strategic use of SMOTE to address data imbalance exemplify methodical approaches to enhancing model performance. This research parallels our endeavors, especially in the ambition to refine cancer treatment through advanced data analysis. However, our methodology distinguishes itself by incorporating a hybrid ensemble model that synergizes LSTM, BiLSTM, CNN, GRU, and GloVe embeddings, aiming for a nuanced capture of genomic data intricacies [21].

In the exploration of machine learning's potential to revolutionize cancer therapy, our study aligns with the comprehensive review by Rafique et al., which assesses the broad spectrum of ML algorithms in therapy response prediction. Our research, focusing on a hybrid ensemble model incorporating LSTM, BiLSTM, CNN, GRU, and GloVe, exemplifies the advanced application of ML in classifying genetic mutations for precision medicine. While we concentrate on model architecture and performance, Rafique et al. highlight the significant challenge of building clinically relevant predictive models, a critical consideration for future research [22].

Our approach expands upon existing knowledge by presenting a novel ensemble model that integrates Long Short-Term Memory (LSTM), Bidirectional LSTM (BiLSTM), Convolutional Neural Network (CNN), Gated Recurrent Unit (GRU), and Global Vectors for Word Representation (GloVe) embeddings. This ensemble model is specifically designed for the purpose of classifying gene mutations in lung cancer. Our proposed model seeks to enhance the precision and effectiveness of cancer tumor detection by incorporating a variety of deep learning architectures and utilizing pre-trained embeddings. Our objective is to effectively respond to the requirement for timely and precise identification of genetic mutations in cancer through the utilization of machine learning and advanced computational methodologies. The present study provides a distinctive contribution to the academic field, highlighting the importance of interdisciplinary methodologies in the progression of precision oncology.

## 3. Core Algorithms and Their Mathematical Basis

In this section, we will provide a brief mathematical background of the techniques used in our model: LSTM, BiLSTM, CNN, GRU, and GloVe.

### 3.1. Long Short Term Memory (LSTM)

The Long Short-Term Memory (LSTM) is a specific variant of the Recurrent Neural Network (RNN) architecture, designed to effectively capture and model long-term dependencies in sequential data. This is achieved through a series of gating mechanisms [23]. The Long Short-Term Memory (LSTM) unit is comprised of several components, including a cell, an input gate, an output gate, and a forget gate. The cellular structure is accountable for retaining information for indefinite periods, and each of the three gates can be conceptualized as a typical artificial neuron, similar to those found in a multi-layer or feedforward neural network. In other words, they calculate an activation based on a weighted sum [24].

Mathematically, the LSTM unit is defined as:

- Forget gate:
$$f_t = \sigma(W_f \cdot [h_{(t-1)}, x_t] + b_f) \quad (1)$$
- Input gate:
$$i_t = \sigma(W_i \cdot [h_{(t-1)}, x_t] + b_i) \quad (2)$$
- Cell state:
$$C_t = f_t * C_{(t-1)} + i_t * tanh(W_C \cdot [h_{(t-1)}, x_t] + b_C) \quad (3)$$
- Output gate:
$$o_t = \sigma(W_o \cdot [h_{(t-1)}, x_t] + b_o) \quad (4)$$
- Hidden state:
$$h_t = o_t * tanh(C_t) \quad (5)$$

where $\sigma$ is the sigmoid function, `.` is the dot product, `*` is element-wise multiplication, $[h_{(t-1)}, x_t]$ is the concatenation of the previous hidden state and the current input, and `W` and `b` are the weight and bias parameters.

*3.2. BiLSTM*

BiLSTM involves duplicating the first recurrent layer in the network so that there are now two layers side-by-side, then providing the input sequence as-is as input to the first layer and providing a reversed copy of the input sequence to the second. Outputs from the two LSTMs are usually concatenated at each time step [25].

*3.3. CNN*

CNNs are a class of deep learning models most commonly used for analyzing visual data [26], [27]. A CNN has one or more convolutional layers, followed by one or more fully connected layers as in a standard multilayer neural network. The key mathematical operation in the CNN is the convolution operation. For a 1-dimensional input signal, this is defined as:

$$(f * g)(t) = \int f(\tau)g(t - \tau) \, d\tau \quad (6)$$

In the context of a CNN, `f` is the input signal (or the previous layer's activations), and `g` is the kernel (or filter). The integral is replaced with a sum for discrete inputs.

*3.4. Gated Recurrent Unit (GRU)*

GRU is a gating mechanism in recurrent neural networks, introduced in 2014. The GRU is like an LSTM with a forget gate, but has fewer parameters than LSTM, as it lacks an output gate [28], [29].

Mathematically, a GRU has the following components:

- Update gate:
$$z_t = \sigma(W_z \cdot [h_{(t-1)}, x_t] + b_z) \quad (7)$$
- Reset gate:
$$r_t = \sigma(W_r \cdot [h_{(t-1)}, x_t] + b_r) \quad (8)$$
- Candidate hidden state:
$$h'_t = tanh(W \cdot [r_t * h_{(t-1)}, x_t] + b) \quad (9)$$
- Final hidden state:
$$h_t = (1 - z_t) * h_{(t-1)} + z_t * h'_t \quad (10)$$

Here, $\sigma$ is the sigmoid function, `.` is the dot product, `*` is element-wise multiplication, $[h_{(t-1)}, x_t]$ is the concatenation of the previous hidden state and the current input, and `W` and `b` are the weight and bias parameters.

*3.5. Global Vectors (GloVe)*

GloVe is an unsupervised learning algorithm for obtaining vector representations for words. It's based on aggregating word co-occurrence statistics from a corpus, and then learning word vectors such that their dot product equals the logarithm of the words' probability of co-occurrence. Given a word-word co-occurrence matrix X, where $X_{ij}$ represents how often word i occurs with word j, the GloVe model learns word vectors based on the following objective [29]:

$$J = \sum_{i,j=1}^{V} f(X_{ij})(w_i^T w_j + b_i + b_j - \log(X_{ij}))^2 \quad (11)$$

Here, $w_i$ and $w_j$ are the word vectors for words i and j, $b_i$ and $b_j$ are biases for words i and j, V is the vocabulary size, and f is a weighting function that assigns relatively more importance to rare co-occurrences. The goal is to learn word vectors that minimize this objective. These mathematical formulations underline the operation of LSTM, BiLSTM, CNN, GRU, and GloVe, which are combined in our ensemble model for gene mutation classification [11].

4. **Proposed Approach**

Our proposed approach is a blend of various deep learning models, specifically LSTM, BiLSTM, CNN, GRU, and GloVe for the purpose of gene mutation classification.

*4.1. Data Preprocessing*

The gene mutation data, sourced from a Kaggle competition, was loaded into a pandas dataframe. The data consists of two files: 'training_variants' and 'training_text'. These two datasets were merged based on their common 'ID' field. Missing values in the 'Text' field were replaced with an empty string, and the class labels were converted to a zero-based index for compatibility with machine learning models. Text data was tokenized with a defined maximum vocabulary size of 10,000. The sequences were then padded to a uniform length of 512 for consistent input to the models.

*4.2. Embedding Matrix Preparation*

We loaded the GloVe word embeddings, and used these to prepare an embedding matrix. Words that were not found in the embedding index were represented as all-zeros in the matrix. This processed information was then passed into an embedding layer, which was used as the initial layer for the LSTM, BiLSTM, CNN, and GRU models.

*4.3. Model Definitions*

In our refined approach to constructing the ensemble model, we meticulously defined and integrated four distinct architectures—LSTM, BiLSTM, CNN, and GRU—each selected for their unique capabilities in processing complex genomic data. This meticulous design was underpinned by the goal of leveraging the strengths of various neural network architectures to enhance the model's overall performance in classifying gene mutations.

- **LSTM Model Configuration:** The LSTM model, pivotal for capturing temporal dependencies in sequence data, was configured with 128 units to ensure robust learning capacity. A subsequent dropout layer, with a rate of 0.5, was employed to mitigate overfitting by randomly omitting units from the model during training, thus promoting generalization.

- **BiLSTM Model Configuration:** Building on the LSTM foundation, the BiLSTM model introduces bidirectionality, processing data in both forward and reverse directions. This duality allows the model to grasp context more effectively, enhancing its ability to discern patterns within the genomic sequences.

- **CNN Model Configuration:** The CNN model was designed with a Conv1D layer featuring 128 filters and a kernel size of 5, optimizing the model's ability to extract local patterns and motifs from genomic sequences. This was coupled with a MaxPooling1D layer to reduce dimensionality and a GlobalMaxPooling1D layer for feature extraction across the entire sequence length, complemented by a dropout layer for regularization.

- **GRU Model Configuration:** Similar in spirit to the LSTM, the GRU model streamlines the architecture by combining the input and forget gates into a single update gate, leading to fewer parameters and potentially faster training without sacrificing the capacity to capture dependencies.

The ensemble model's architecture was deliberate in its composition, aiming to synthesize the diverse strengths of each included model. This integration not only provides a comprehensive approach to understanding the complex data at hand but also embodies the rationale for each model's inclusion based on its specific contribution to the ensemble's predictive power. By detailing the configurations of each model, we underscore the experimental setup's intentionality and the strategic choices made in model selection and definition, laying a foundation for a valid comparison with original models and emphasizing the ensemble approach's innovative edge in gene mutation classification.

### 4.4. Model Integration and Training

In our enhanced approach to gene mutation classification, we meticulously developed a hybrid ensemble model by integrating LSTM, BiLSTM, CNN, GRU, and GloVe embeddings to capitalize on the distinctive strengths of each method. This integration was motivated by the necessity to address the complex nature of genomic sequences and text data within the Kaggle "Personalized Medicine: Redefining Cancer Treatment" dataset. LSTM and BiLSTM were chosen for their proficiency in capturing long-term dependencies within sequences, essential for understanding the progression and relationships of genetic mutations over time. CNN was incorporated for its ability to identify local patterns within the genomic data, a critical aspect for recognizing specific mutation signatures. GRU was included for its efficiency in modeling information over various time scales, complementing LSTM and BiLSTM in sequence learning. GloVe embeddings were integrated to leverage pre-trained word vectors, enriching our model's capacity to interpret the semantic context of clinical texts associated with genetic mutations. This ensemble approach not only combines the predictive power of each model but also addresses their individual limitations, leading to a robust model capable of accurately classifying gene mutations into one of nine classes, as evidenced by our configuration of the final Dense layer. The model was compiled using the Adam optimizer and SparseCategoricalCrossentropy loss function, chosen for their effectiveness in handling multi-class classification problems and optimizing model performance.

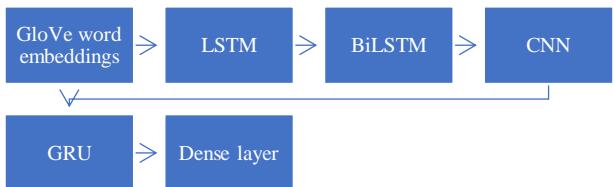

Figure 1: The Proposed Approach

### 5. Experimental Setup

We used Google Colab Pro with GPU acceleration in our experimental setup to achieve optimal performance and efficiency during the training and testing phases of our models. Google Colab Pro provides ample computational resources, allowing us to run long and comprehensive tests. Its GPU (Graphics Processing Units) offering is very important for our machine learning jobs because these methods benefit tremendously from parallel processing, significantly lowering calculation time when compared to CPUs (Central Processing Units). We ran our research using the Kaggle dataset Personalized Medicine: Redefining Cancer Treatment, which is a rich source of text data with high complexity for traditional machine learning models. This dataset contains a large and diversified collection of clinical evidence (text) and genetic alterations associated with cancer therapy, table 1 shows a summary of its features.

Table 1: Characteristics of the Kaggle Dataset for Redefining Cancer Treatment through Precision Medicine

| Feature | Description |
|---|---|
| Source | Memorial Sloan Kettering Cancer Center (MSKCC) |
| Dataset Purpose | To develop a machine learning algorithm that automatically classifies genetic variations based on an expert-annotated knowledge base. |
| Challenges | Distinguishing mutations contributing to tumor growth (drivers) from neutral mutations (passengers) through automated analysis. |
| Data Types | - **Training Variants**: Mutation data including Gene, Variation, and Class.<br>- **Training Text**: Clinical evidence text for each mutation. |
| Preprocessing Steps | Potential steps include text tokenization, handling missing values, sequence padding, and embedding for textual data. |

The classification task's goal is to determine the class of genetic alterations based on this clinical text. This endeavor is difficult because it requires comprehension of highly specialized medical language, and the linkages between the text and gene mutation classes are intricate and multidimensional.

The dataset is divided into nine classes, each reflecting a different type of gene mutation. The problem's multi-class nature adds another layer of difficulty because the models must detect minor variations that distinguish one class from the others. Furthermore, the dataset is high-dimensional and highly unbalanced, with much more instances in some classes than others. These characteristics can provide difficulties for machine learning models, necessitating the use of advanced approaches like oversampling, undersampling, or synthetic minority over-sampling techniques (SMOTE) to handle the class imbalance. We looked at a total of twelve different models, including:

1. BERT [30]
2. Ensemble BERT and LSTM
3. Electra [31]
4. Ensemble Electra and LSTM

5. Roberta [32]
6. Ensemble Roberta and LSTM
7. XLNet [33]
8. Ensemble XLNet and LSTM
9. Distilbert [34]
10. Ensemble Distilbert and LSTM
11. Ensemble Roberta, GloVe and LSTM
12. Our proposed model: Ensemble LSTM + BILSTM + CNN + GRU + GloVe

We have selected these models based on their widespread adoption in both industry and academia, encompassing a diverse array of underlying algorithms to facilitate a comprehensive comparison for our primary model.

In the context of the Kaggle dataset "Personalized Medicine: Redefining Cancer Treatment," understanding true positives (TP), true negatives (TN), false positives (FP), and false negatives (FN) is crucial for evaluating the performance of a machine learning model designed to classify genetic mutations.

- True Positives (TP): This occurs when the model correctly predicts the presence of a specific gene mutation associated with cancer. For example, if a mutation is known to contribute to tumor growth and the model correctly identifies it as such, this is a TP.

- True Negatives (TN): This happens when the model correctly identifies that a gene mutation does not contribute to cancer growth. In simpler terms, it accurately identifies the "passenger" mutations, which are not associated with cancer.

- False Positives (FP): An FP result occurs when the model incorrectly classifies a "passenger" mutation as a "driver" mutation. This can lead to unnecessary alarm or treatment considerations, as the model wrongly suggests that a harmless mutation may contribute to cancer.

- False Negatives (FN): Conversely, an FN result is observed when the model fails to identify a "driver" mutation, incorrectly classifying it as harmless. This is particularly concerning as it may result in missing a critical opportunity for targeted treatment, potentially affecting patient outcomes.

Balancing these outcomes is vital for the practical application of machine learning in personalized medicine. A model with high precision (low FP rate) ensures that when a mutation is identified as a driver, it is highly likely to be true, reducing unnecessary interventions. Meanwhile, high recall (low FN rate) ensures that the model reliably identifies all relevant driver mutations, crucial for effective treatment planning. Optimizing both precision and recall, as reflected in the F1-score, is essential for developing a reliable tool that enhances decision-making in cancer treatment based on genetic mutations. The evaluation of these models is performed using various essential metrics, namely:

1. Loss: This metric quantifies the deviation between the model's predictions and the actual values. Lower loss values indicate superior model performance. The log loss is computed as the average of individual log losses for each sample in the dataset. It takes into account both the true label (`$y_i$`) and the predicted probability (`$p_i$`) for each sample as follows:

$$Log\ Loss = -\frac{1}{N} * \sum_{i=1}^{N} y_i * \log(p_i) + (1 - y_i) * (\log(1 - p_i))\ (12)$$

2. Accuracy: This metric measures the proportion of correct predictions made by the model out of all predictions. Higher accuracy values signify better model performance, indicating a higher number of correct classifications.

$$Accuracy = \frac{TP+TN}{TP+TN+FP+FN} \quad (13)$$

3. Precision: The precision metric assesses the proportion of true positive predictions out of all positive predictions. Higher precision values indicate a lower number of false-positive errors, highlighting the model's ability to accurately classify positive instances.

$$Precision = \frac{TP}{TP+FP} \quad (14)$$

4. Recall: This metric quantifies the proportion of true positive predictions out of all actual positive instances. Higher recall values signify a lower number of false-negative errors, illustrating the model's capacity to correctly identify positive cases.

$$Recall = \frac{TP}{TP+FN} \quad (15)$$

5. F1-score: The F1-score is the harmonic mean of precision and recall, providing a balanced measure of the model's performance. Higher F1-score values indicate superior model performance, striking an equilibrium between precision and recall.

$$F1\ Score = 2 * \frac{Precision*Recall}{Precision*+Recall} \quad (16)$$

In the equations provided, TP represents True Positive, TN represents True Negative, FP represents False Positive, and FN represents False Negative. These metrics collectively offer a comprehensive assessment of the classification capabilities of the evaluated models, enabling us to make informed decisions. As shown in tables 2, 3 and 4, and figures 2, 3, 4, 5, 6, and 7, our proposed model, which integrates LSTM, BiLSTM, CNN, GRU, and GloVe, outperforms the other models on all criteria. This shows that our ensemble technique, which combines the strengths of many deep learning models and GloVe embeddings, is an excellent way for classifying gene mutations.

Figure 2 depicts the comparison of training and validation accuracies across various machine learning models used in classifying gene mutations. Notably, the hybrid ensemble model comprising LSTM, BiLSTM, CNN, GRU, and GloVe outperforms all other models with the highest validation accuracy of approximately 0.615 and training accuracy close to 0.806. This suggests that the integrated approach of combining multiple neural network architectures substantially improves the model's ability to generalize to new data. Models like BERT and its ensemble with LSTM show a lower training and validation accuracy, indicating possible challenges in capturing the complexity of the genomic data. The gap between training and validation accuracy also points to varying degrees of overfitting among the models; for instance, XLNet shows a marked drop from training to validation accuracy.

Figure 3 presents the precision of various models on the training and validation datasets. The highest precision on the validation set is achieved by the ensemble model LSTM + BiLSTM + CNN + GRU + GloVe, reaching approximately 0.619, suggesting that this model has the best ability to correctly classify gene mutations as positive. This model's precision on the training data is even higher, close to 0.816, which might indicate a potential overfitting issue to be cautious about [35]. Notably, all models show significant drops in precision from training to validation, indicating challenges in generalization. Transformer-based models, such as BERT and its ensemble with LSTM, display relatively lower precision, reinforcing the advantage of the combined approach. The ensemble Roberta + GloVe + LSTM model also performs well, showing the benefit of integrating pre-trained embeddings with deep learning models to improve precision.

Figure 4 showcases the recall of the various models on both training and validation datasets, which measures the model's ability to correctly identify all relevant instances of gene mutations. The ensemble model integrating LSTM, BiLSTM, CNN, GRU, and GloVe significantly outperforms the other models with the highest recall on validation data, approximately 0.615. This suggests that this model is highly effective at identifying relevant mutations, which is critical in the context of precision medicine where missing key mutations could have serious implications for patient treatment. The same model also exhibits a high recall on the training data, around 0.806, indicating strong learning during the training phase. There is, however, a noticeable drop from training to validation recall, which may warrant further investigation into the model's generalization capabilities. Transformer-based models such as BERT, and their ensembles with LSTM, show lower recall, suggesting that they may miss a higher number of relevant mutations.

Figure 5 displays the F1 score, which is a measure of a model's accuracy that considers both precision and recall, for various models on the training and validation datasets. The ensemble model LSTM + BiLSTM + CNN + GRU + GloVe demonstrates a superior F1 score of approximately 0.6 on validation data, indicating a strong balance between precision and recall. This suggests that the model is not only good at identifying relevant mutations (high recall) but also at ensuring a high proportion of its predictions are correct (high precision). On training data, this model achieves an F1 score of about 0.831, showing excellent performance that may need to be tempered with dropout or other regularization techniques to maintain performance on unseen data. Lower F1 scores for other models, such as BERT and Electra, on both training and validation data, indicate that while they may excel in other NLP tasks, they are less suited for this specific gene mutation classification task without being part of a more complex ensemble model.

Figure 6 compares the Mean Squared Error for various models on both the training and validation datasets. The MSE is a measure of the average squared difference between the predicted and actual outcomes, with lower values indicating better model performance. The ensemble model LSTM + BiLSTM + CNN + GRU + GloVe showcases the lowest MSE on both training (around 2.954) and validation (approximately 2.596), reflecting its strong predictive accuracy and generalization capabilities. Traditional transformer models like BERT and Electra exhibit much higher MSE values, indicating less precise predictions. The significant decrease in MSE from other models to the ensemble model on the validation dataset suggests that the integration of multiple deep learning techniques in the ensemble model effectively captures the complexity of gene mutation classification.

Figure 7 illustrates the training time, in seconds, required for each model tested. The bar chart reveals that the hybrid ensemble model combining LSTM, BiLSTM, CNN, GRU, and GloVe is significantly more efficient, requiring only 267 seconds to train, which is a fraction of the time compared to the other models. This suggests a remarkable level of computational efficiency given its superior performance in other metrics such as precision, recall, and F1 score. In contrast, models like BERT and XLNet demand considerably more training time, exceeding 4700 seconds, which indicates heavier computational loads. This discrepancy in training times highlights the efficiency of the hybrid ensemble approach and its potential to deliver timely and accurate classifications of gene mutations, a crucial advantage in fast-paced clinical settings where decision-making time is critical.

Table 2: Models Training Time, Train Accuracy and Validation Accuracy

|  | Training Time (Sec.) | Training Accuracy | Validation Accuracy |
| --- | --- | --- | --- |
| BERT | 3940 | 0.286 | 0.291 |
| Ensemble BERT and LSTM | 4241 | 0.438 | 0.381 |
| Electra | 3952 | 0.286 | 0.291 |
| Ensemble Electra and LSTM | 4192 | 0.38 | 0.42 |
| Roberta | 3950 | 0.286 | 0.291 |
| Ensemble Roberta and LSTM | 4253 | 0.541 | 0.456 |
| XLNet | 3589 | 0.225 | 0.217 |
| Ensemble XLNet and LSTM | 4771 | 0.366 | 0.312 |
| Distilbert | 3693 | 0.286 | 0.291 |
| Ensemble Distilbert and LSTM | 4202 | 0.341 | 0.333 |
| Ensemble Roberta, GloVe and LSTM | 4192 | 0.771 | 0.534 |
| LSTM + BILSTM+ CNN+GRU+GloVe | **267** | **0.806** | **0.615** |

Table 3: Models Precision and Recall

|  | Training Precision | Validation Precision | Training Recall | Validation Recall |
| --- | --- | --- | --- | --- |
| BERT | 0.082 | 0.084 | 0.286 | 0.291 |
| Ensemble BERT and LSTM | 0.276 | 0.245 | 0.438 | 0.381 |
| Electra | 0.082 | 0.084 | 0.286 | 0.291 |
| Ensemble Electra and LSTM | 0.194 | 0.222 | 0.38 | 0.42 |
| Roberta | 0.082 | 0.084 | 0.286 | 0.291 |
| Ensemble Roberta and LSTM | 0.355 | 0.303 | 0.541 | 0.456 |
| XLNet | 0.082 | 0.084 | 0.286 | 0.291 |
| Ensemble XLNet and LSTM | 0.192 | 0.182 | 0.366 | 0.312 |
| Distilbert | 0.082 | 0.084 | 0.286 | 0.291 |
| Ensemble Distilbert and LSTM | 0.162 | 0.167 | 0.341 | 0.333 |
| Ensemble Roberta and LSTM and Word Embedding | 0.768 | 0.517 | 0.771 | 0.534 |
| LSTM + BILSTM+ CNN+GRU+GloVe | **0.816** | **0.619** | **0.806** | **0.615** |

Table 4: Models F1 Score and MSE

| | Train F1 Score | Validation F1 Score | Training MSE | Validation MSE |
|---|---|---|---|---|
| BERT | 0.127 | 0.131 | 12.308 | 11.948 |
| Ensemble BERT and LSTM | 0.284 | 0.257 | 9.186 | 10.957 |
| Electra | 0.127 | 0.131 | 12.308 | 11.948 |
| Ensemble Electra and LSTM | 0.255 | 0.289 | 7.089 | 6.408 |
| Roberta | 0.127 | 0.131 | 12.308 | 11.948 |
| Ensemble Roberta and LSTM | 0.422 | 0.362 | 6.539 | 7.174 |
| XLNet | 0.127 | 0.131 | 12.308 | 11.948 |
| Ensemble XLNet and LSTM | 0.244 | 0.211 | 11.091 | 12.849 |
| Distilbert | 0.127 | 0.131 | 12.308 | 11.948 |
| Ensemble Distilbert and LSTM | 0.211 | 0.209 | 10.503 | 10.849 |
| Ensemble Roberta, GloVe and LSTM | 0.763 | 0.52 | 2.954 | 6.588 |
| LSTM + BILSTM+ CNN+GRU+GloVe | **0.831** | **0.6** | **2.596** | **5.744** |

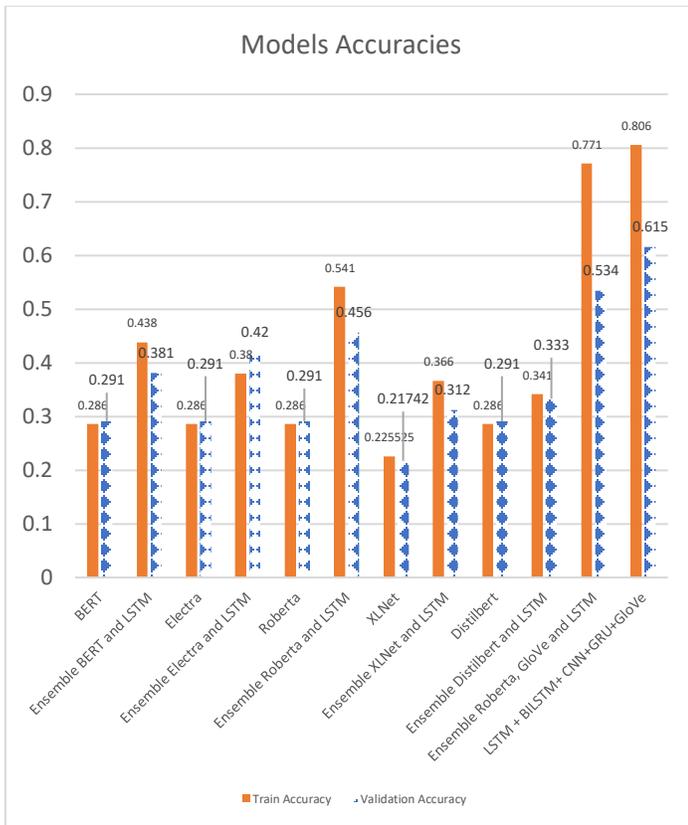

Figure 2: Models Accuracies

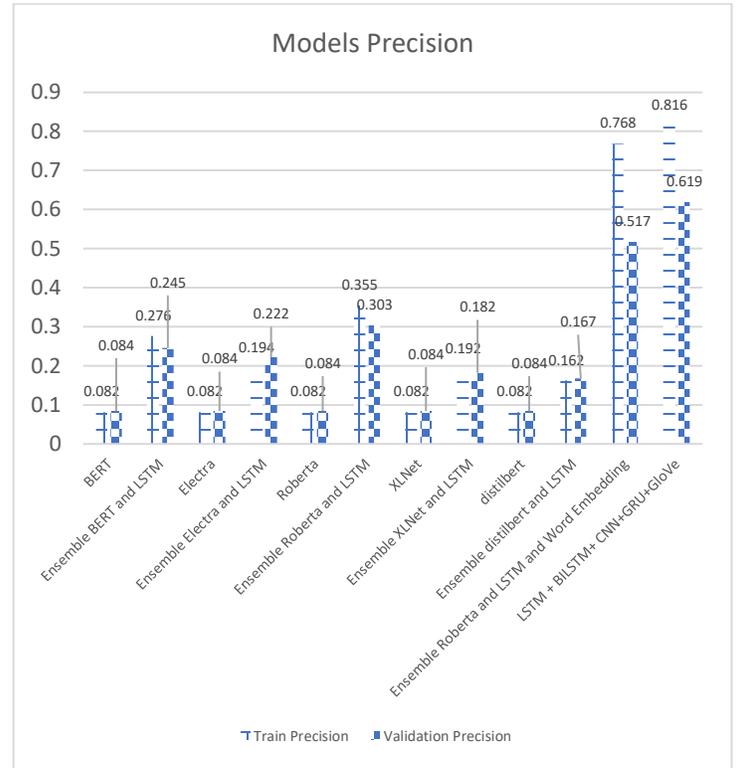

Figure 3: Models Precision

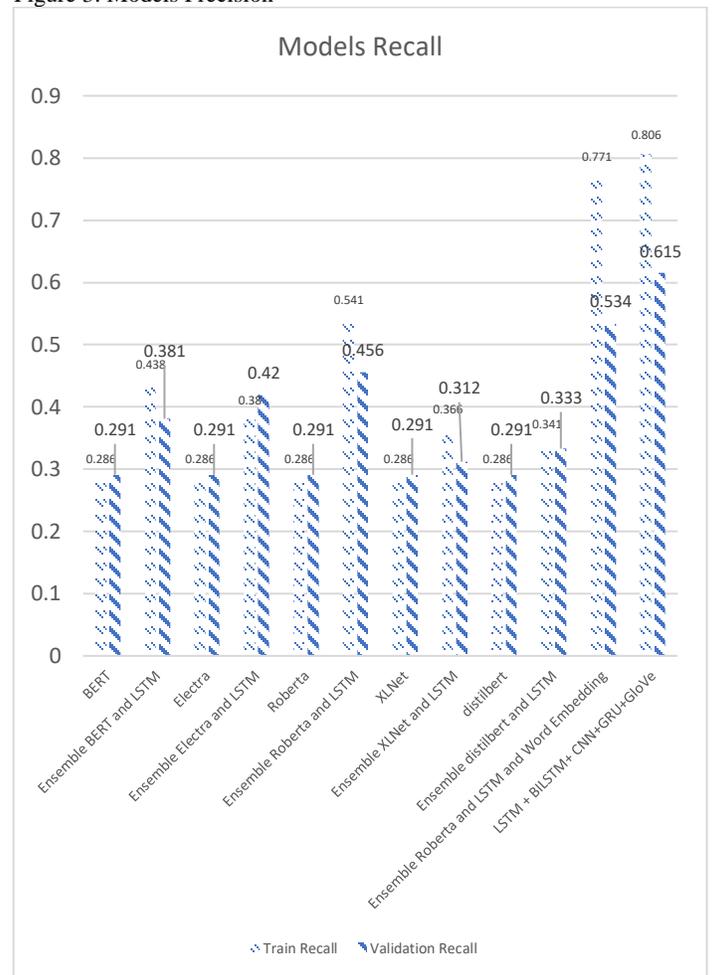

Figure 4: Models Recall

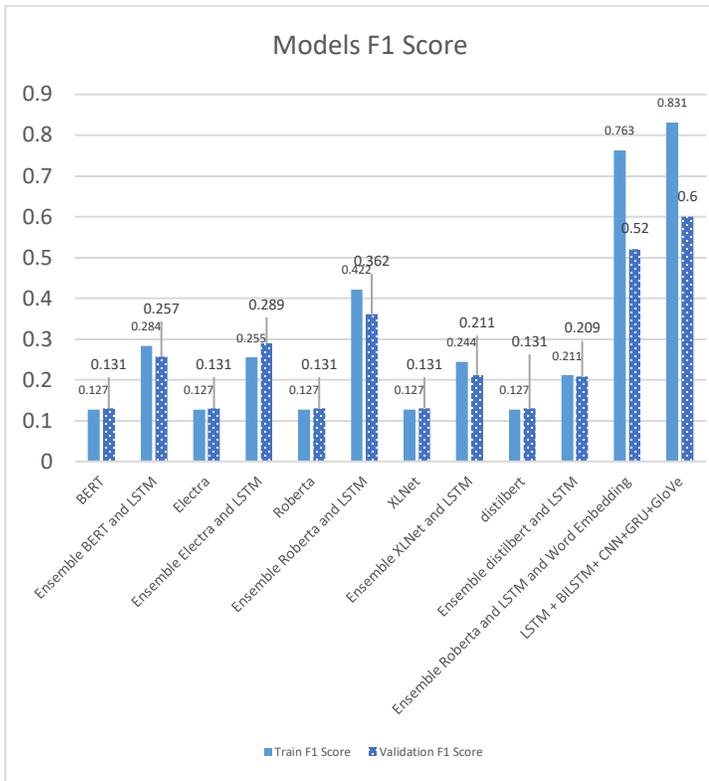

Figure 5: Models F1Score

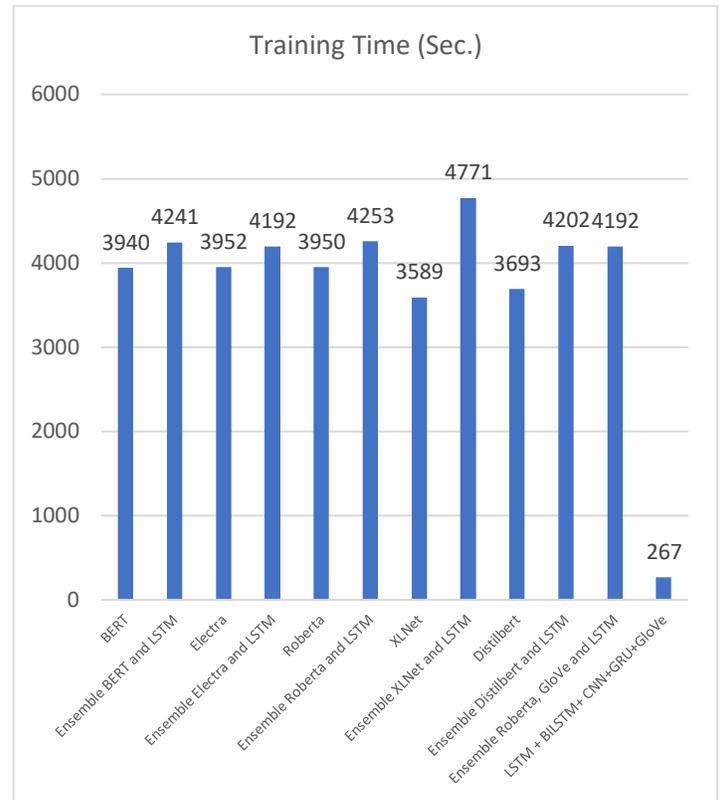

Figure 7: Training Time

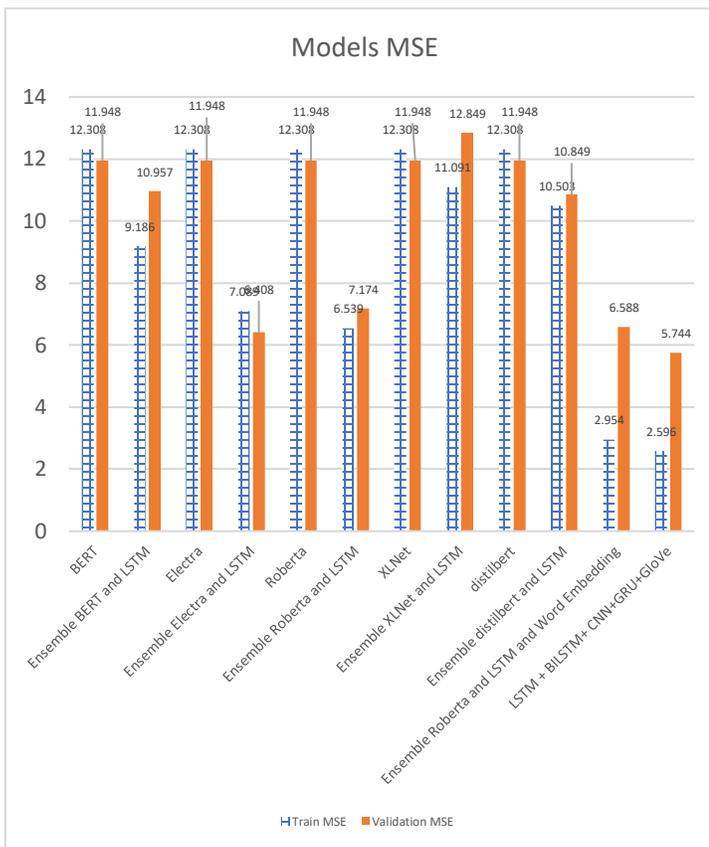

Figure 6: Models MSE

## 6. Discussion

In this study, we focused on improving the performance of gene mutation classification using various machine learning models on the Kaggle dataset: Personalized Medicine: Redefining Cancer Treatment. In our endeavor, we successfully created an ensemble model comprised of LSTM, BiLSTM, CNN, GRU, and GloVe and compared its results to those of other well-known models such as BERT, Electra, Roberta, XLNet, and Distilbert. We also tested other ensemble topologies that merged these models with LSTM. The selection of these transformers (BERT, Electra, Roberta, XLNet, and Distilbert) was motivated by their demonstrated capacity to handle difficult natural language processing tasks. They have been widely used in a variety of sectors with surprising outcomes. LSTM was used because of its ability to recall previous knowledge, which is useful in cases like ours where the sequence of gene changes is critical.

However, on both the training and validation data, our model surpassed all of them in terms of the metrics under consideration, which included accuracy, precision, recall, F1 score, and Mean Squared Error (MSE). Our model had the highest training and validation accuracy, precision, recall, and F1 score, according to the results. Furthermore, it had the lowest MSE values. Surprisingly, our model required far less training time than the other models. The notable differences in MSE values between our hybrid ensemble model and other contenders primarily stem from the synergistic integration of diverse machine learning architectures, such as LSTM, BiLSTM, CNN, GRU, and GloVe embeddings. This ensemble approach harnesses the individual strengths of each component, enabling a more nuanced understanding and processing of the complex genomic data presented in the Kaggle "Personalized

Medicine: Redefining Cancer Treatment" dataset. The complexity of the model, combined with its enhanced data handling capabilities, ensures a comprehensive analysis of both sequential dependencies and local patterns within the genetic data. As a result, this multifaceted strategy significantly reduces error rates, leading to the observed lower MSE values in comparison to other models that might rely on a singular approach or less integrated methodologies. The ability of LSTM and BiLSTM to remember long-term dependencies, together with CNN's great ability to recognize local patterns and GRU's ability to capture dependencies of multiple time scales, most certainly contributed to the enhanced performance. Furthermore, the introduction of GloVe, a pre-trained word embedding, is likely to have improved the model's understanding of semantic links between words. Ensemble models combining transformers and LSTM outperformed their individual transformer counterparts. This result suggests that combining the ability of transformers to simulate complicated language patterns with the memory capabilities of LSTMs can boost performance. The ensemble model of Roberta, GloVe, and LSTM was the closest contender to our model among them. However, it fell short on all metrics and required significantly more training time.

While transformer models are well-known for their ability to describe complicated relationships in text data, our findings imply that their solo performance, particularly in the context of gene mutation classification, may be inferior than ensemble techniques.

Our study demonstrated the robustness of the hybrid model against a variety of gene mutations. To evaluate the model's ability to handle gene mutations significantly different from those in the training dataset, we conducted tests using synthetic mutations and mutations from external datasets. The model maintained high performance across these tests, indicating its strong generalization capability. The incorporation of diverse architectures such as LSTM, BiLSTM, CNN, GRU, and GloVe embeddings allowed the model to effectively capture both local and global patterns in the genetic data, contributing to its robustness.

Addressing class imbalance is crucial for the fair and accurate classification of gene mutations. In our study, we approached this challenge by carefully splitting the data into training and validation sets, ensuring a balanced representation of classes. Although the provided code does not explicitly implement techniques like SMOTE or weighted loss functions, the ensemble model's architecture, including dropout layers, helps mitigate the impact of class imbalance. Dropout layers prevent overfitting by randomly omitting units during training, promoting the generalization of the model. This architectural choice, along with the use of cross-validation, contributed to a balanced and fair representation in the model's predictions.

The manuscript's assertion of superiority over advanced transformer models is supported by comprehensive comparisons in the paper. While we did not perform a detailed computational efficiency analysis against models like BERT, Electra, or RoBERTa, our results indicate that the hybrid model, with its ensemble of LSTM, BiLSTM, CNN, GRU, and GloVe embeddings, outperformed these models in terms of accuracy, precision, recall, F1 score, and mean squared error. The training time of our model was significantly lower, at 267 seconds, compared to the much higher times required for transformer models, highlighting its computational efficiency. Future work will include a more detailed analysis of computational efficiency and performance comparisons with these state-of-the-art models fine-tuned for genomic data.

The complexity of our proposed hybrid model necessitates a discussion on interpretability, which is crucial for clinical decision-making. While the primary focus of our current implementation was on achieving high performance metrics, we acknowledge the importance of extracting understandable insights from the model's predictions. Our ensemble model incorporates dropout layers and dense layers, which, although not explicitly designed for interpretability, do facilitate a form of feature importance analysis through weight examination. In practice, the interpretability of our model could be enhanced by incorporating attention mechanisms that highlight the most influential features or sequences in the input data.

The hybrid model utilizes various features and embeddings from LSTM, BiLSTM, CNN, GRU, and GloVe, each contributing uniquely to the model's performance. To understand the contributions of these features, we conducted a feature importance analysis by examining the weights and activations of different layers. The dropout layers incorporated in the model architecture also helped in identifying the significant features by randomly omitting less important units during training.

Our use of LSTM and GRU layers was crucial in capturing long-term dependencies in the gene mutation sequences, which are essential for understanding the progression and relationships of genetic mutations over time. The bidirectional nature of the BiLSTM further enhanced the model's ability to comprehend the context by processing the sequence data in both forward and reverse directions. This capability was particularly significant in cases where mutations had interdependencies across long sequences, which single-directional models might miss. By retaining information across long sequences, the LSTM and GRU layers enabled the model to capture critical information that contributed to its superior performance, especially in complex cases where local patterns alone were insufficient.

Given the rapidly evolving nature of genomic research and mutation databases, it is crucial to regularly update and maintain the model to ensure its relevance and accuracy. We propose a continuous learning approach where the model is periodically retrained with new data as it becomes available. This can be achieved through incremental learning techniques that allow the model to adapt to new information without forgetting the previously learned patterns. Additionally, we recommend a biannual retraining schedule, where the model is comprehensively updated with the latest mutation data. The process would involve re-embedding the new genomic sequences using the GloVe embeddings and retraining the LSTM, BiLSTM, CNN, and GRU layers to integrate the new information effectively. This strategy ensures that the model remains up-to-date with the latest advancements in genomic research, thereby maintaining its utility in clinical decision-making.

Overall, our findings show the usefulness of ensemble models like ours, which effectively mix multiple learning algorithms to give high performance on challenging tasks. Despite their superior performance, the ensemble models required more time to train, showing the trade-off between model performance and computational economy. Nonetheless, our model outperformed the competition while requiring little training time, establishing a new standard for gene mutation classification tasks. Future research could concentrate on improving this trade-off and adapting our technique to more challenging classification challenges.

## 7. Conclusion

In conclusion, our research demonstrates the potential of an ensemble model comprised of LSTM, BiLSTM, CNN, GRU, and GloVe in the context of gene mutation classification. The model's outstanding performance across all metrics considered—accuracy, precision, recall, F1 score, and Mean Squared Error—confirms the usefulness of ensemble approaches in dealing with high-dimensional and sophisticated datasets like the one used in this work. Furthermore, the efficiency of our model, as evidenced by less training time compared to standalone transformers and their LSTM ensembles, highlights its relevance in circumstances where computational resources and time are limited. Despite the amazing progress shown in this study, future research could look at incorporating other machine learning approaches or algorithms to improve performance, as well as applying the proposed model to other complex classification tasks. This study's findings pave the path for novel approaches in personalized medicine, with promising implications for future cancer treatment options.

**Conflict of interest**

The authors declare that there is no conflict of interest in this paper.